\documentclass[aps,twocolumn,showpacs,floats,superscriptaddress]{revtex4}
\usepackage{graphicx}
\usepackage{dcolumn}
\usepackage{bm}
\usepackage{color}

\begin{document}

\author{S. Pilati}
\affiliation{Theoretische Physik, ETH Zurich, CH-8093 Zurich, Switzerland}
\author{G. Bertaina}
\affiliation{Dipartimento di Fisica, Universit\`a di Trento and CNR-INO BEC Center, I-38050 Povo, Trento, Italy}
\author{S. Giorgini}
\affiliation{Dipartimento di Fisica, Universit\`a di Trento and CNR-INO BEC Center, I-38050 Povo, Trento, Italy}
\author{M. Troyer}
\affiliation{Theoretische Physik, ETH Zurich, CH-8093 Zurich, Switzerland}

\title{Itinerant ferromagnetism of a repulsive atomic Fermi gas: \\ 
a quantum Monte Carlo study}

\begin{abstract} 
We investigate the phase diagram of a two-component repulsive Fermi gas at $T=0$ by means of quantum Monte Carlo simulations. For a given value of the positive $s$-wave scattering length, both purely repulsive and purely attractive model potentials are considered in order to analyze the limits of the universal regime where the details of interatomic forces can be neglected. The equation of state of both balanced and unbalanced systems is calculated as a function of the interaction strength and the critical density for the onset of ferromagnetism is determined. The energy per particle of the strongly polarized gas is calculated and parametrized in terms of the physical properties of repulsive polarons, which are relevant  for the stability of the fully magnetized ferromagnetic state. Finally, we analyze the phase diagram in the polarization/interaction plane under the assumption that only phases with homogeneous magnetization can be produced.   
\end{abstract}

\pacs{05.30.Fk, 03.75.Hh, 75.20.Ck}
\maketitle 

Over the past decade there has been substantial progress in the  experimental realization of quantum degenerate atomic Fermi gases. A major part of the activity carried out so far was devoted to the investigation of the role of {\em attractive} interactions, with special emphasis on the onset of pairing and superfluidity in the vicinity of a Feshbach resonance as well as in the presence of spin imbalance~\cite{RMP}. More recently attention was drawn to {\em repulsive} interactions and the onset of magnetic behavior. This topic is particularly important in optical lattices because of its connection with the repulsive Hubbard model, a fundamental paradigm of condensed matter physics with still many unanswered questions~\cite{Georges07}, but also for continuous systems where a major recent achievement has been the observation of itinerant ferromagnetism induced by repulsive forces in a two-component Fermi gas~\cite{Jo09}. This experiment realizes the Stoner model, a textbook Hamiltonian that aims to describe itinerant ferromagnetism in an electron gas with screened Coulomb interaction~\cite{Stoner33}.       

On the theoretical side there have been a number of papers addressing the problem of stability of a repulsive two-component Fermi gas~\cite{Houbiers97} and of phase separation in harmonic trapped configurations within the local density approximation~\cite{LDA}. These studies are based on a simple mean-field description of interaction effects that is valid to linear order in the scattering length. In homogeneous systems at $T=0$ they predict a second order phase transition to a magnetized state if the interaction strength is larger than the critical value $k_Fa>\pi/2$, where $a$ is the $s$-wave scattering length and $k_F=(3\pi^2n)^{1/3}$ is the Fermi wave vector in terms of the total particle density of the gas $n=n_\uparrow+n_\downarrow$. An extension of this approach that includes next order corrections to the interaction energy was developed in Ref.~\cite{Duine05} and predicts a smaller value of the critical density ($k_Fa>1.054$), as well as a discontinuous jump in the magnetization. Low-energy theories of itinerant fermions also predict a first-order transition~\cite{Belitz99}. A recent non-perturbative quantum Monte Carlo calculation, instead, suggests the existence of a textured magnetic phase at the border of the ferromagnetic transition and yields the value $k_Fa\simeq0.8$ for the critical density~\cite{Conduit09}. On the other hand, the existence of a ferromagnetic transition has been questioned in Ref.~\cite{Zhai09} by arguing that nonmagnetic states with strong short-ranged repulsive correlations could be energetically favorable compared to ferromagnetic ones.

\begin{figure}
\begin{center}
\includegraphics[width=7.0cm]{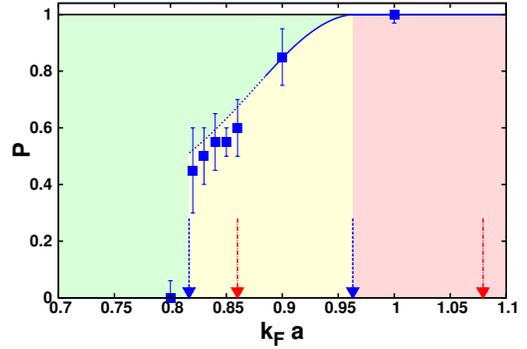}
\caption{(color online). Phase diagram of the HS gas in the interaction/polarization plane. The green region corresponds to the homogeneous phase. The other regions correspond to phase separated states with partially polarized domains (yellow) and fully ferromagnetic domains (pink). The (blue) symbols correspond to the minimum of the curve $E(P)$ and the solid/dashed line is the phase boundary determined from the equilibrium condition for pressure and chemical potentials. The blue and red arrows indicate the critical densities where $\chi$ diverges and full ferromagnetism sets in, respectively for the HS and SW potential.}
\label{fig4}
\end{center}
\end{figure}

Various important issues concerning the regime of strong repulsion are still open. In this Letter we provide answer to some of them, in particular: i) we calculate the equation of state of the Fermi gas using different potentials to determine the regime of interaction strength $k_Fa$ where universality in terms of just the $s$-wave scattering length~$a$ is lost and other parameters of the interatomic potential become relevant;
ii) we study population imbalanced configurations and show that the equation of state of the highly polarized gas can be described in terms of a Fermi liquid of polarons, similarly to the well-established case of attractive interactions~\cite{Polaron1,Polaron2};
iii) we investigate the onset of ferromagnetism and show how the physics of polarons is related to the stability of the ferromagnetic states;
iv) we establish the phase diagram in the polarization/interaction plane where we determine the borders of three phases~\cite{Note1}: a uniformly polarized phase and mixed phases consisting of partially or fully polarized domains, as shown in Fig.~\ref{fig4}.

To address these issues we perform quantum Monte Carlo (QMC) simulations of Fermi systems characterized by a positive scattering length $a>0$, interacting through either purely repulsive or purely attractive forces.
The Hamiltonian of the Fermi gas is 
\begin{equation}
H=-\frac{\hbar^2}{2m}\left( \sum_{i=1}^{N_\uparrow}\nabla^2_i + \sum_{i^\prime=1}^{N_\downarrow}\nabla^2_{i^\prime}\right)
+\sum_{i,i^\prime}V(r_{ii^\prime}) \;,
\label{hamiltonian}
\end{equation}   
where $m$ denotes the mass of the atoms, $i,j,...$ and $i^\prime,j^\prime,...$ label, respectively, spin-up and spin-down
fermions with $N_\uparrow+N_\downarrow=N$, $N$ being the total number of atoms. We model the interspecies interatomic 
interactions using three different potentials: i) a {\it hard sphere} (HS) potential, $V(r)=+\infty$ if $r<a$ and zero otherwise, ii) a repulsive {\it soft sphere} (SS) potential, $V(r)=V_0$ if $r<R_0$ and zero otherwise and iii) an attractive {\it square well} (SW) potential, $V(r)=-V_0$ if $r<R_0$ and zero otherwise ($V_0>0$). The $s$-wave scattering length $a$ coincides with the range of the potential in the HS case and can readily be determined from the range $R_0$ and the strength $V_0$ in the other two cases. For the SS potential $a$ is always smaller than the range $R_0$ and we fix the height $V_0$ such that $a=R_0/2$. In the case of the SW potential instead, the scattering length diverges to $\pm\infty$ every time a new bound state enters the well: we fix the range such that $nR_0^3=10^{-6}$ in terms of the particle density  $n$ and the depth $V_0$ takes values corresponding to the positive branch of $a$ with a single bound state. The short-range SW potential provides a realistic description of interatomic forces in ultracold atoms.

In the case of purely repulsive interactions we use the fixed-node diffusion Monte Carlo (FN-DMC) method. This variational method yields an upper bound for the ground-state energy of the gas, sampling the lowest-energy wave function whose many-body nodal surface is the same as that of a trial wave function $\psi_T$.  FN-DMC can give the exact ground-state energy, provided one knows the exact nodal surface, and in general the energies have been found to be highly accurate even if nodes are only approximate (for more details see {\it e.g.} \cite{Reynolds82}). Our trial wave function is of the Jastrow-Slater form
\begin{equation}
\psi_T({\bf R})=\prod_{i,i^\prime}f(r_{ii^\prime}) D_\uparrow(N_\uparrow) D_\downarrow(N_\downarrow) \;,
\label{psiT}
\end{equation}
where ${\bf R}=({\bf r}_1,..., {\bf r}_N)$ is the spatial configuration vector of the $N$ particles and $D_{\uparrow(\downarrow)}$ denotes the Slater determinant of plane waves in a cubic box of size $L$ with periodic boundary conditions, accommodating the $N_{\uparrow(\downarrow)}$  particles with up (down) spin. The Jastrow correlation term $f(r)$ is obtained from the solution of the two-body scattering problem with the potential $V(r)$, satisfying the boundary condition on its derivative $f^\prime(r=L/2)=0$. Since $f(r)>0$, the many-body nodal surface results only from the antisymmetric character of $\psi_T$ and  coincides with that of a non interacting gas, thus correctly reproducing this limit. Another advantage is that the trial function (\ref{psiT}) can be used to simulate both unpolarized ($N_\uparrow=N_\downarrow$) and polarized ($N_\uparrow>N_\downarrow$) configurations.    

Attractive interactions, as modeled by the SW potential, are more delicate, because of the presence of bound states (molecules) in the true ground state. The atomic Fermi gas of interest here is a meta-stable state consisting of unbound fermionic atoms and no dimers or other bound molecules. Since we are interested in a metastable excited state we cannot use the FN-DMC method 
and resort to the variational Monte Carlo (VMC) calculation that provides a stable estimate of the energy $E_{\text{VMC}}=\langle\psi_T|H|\psi_T\rangle/\langle\psi_T|\psi_T\rangle$. 
The absence of molecular bound states can be readily implemented for two particles by choosing the Jastrow correlation term $f(r)$ to be the scattering solution of the SW potential corresponding to positive energy, which by construction is orthogonal to the bound molecule. A many-body wave function is then constructed using Eq.~(\ref{psiT}) with this choice of $f(r)$, while $f^\prime(r=L/2)=0$ takes care of periodic boundary conditions similarly to the purely repulsive case. For small scattering energies, corresponding to large values of $L$, the Jastrow term $f$ changes sign at $r=a$. The larger the size of the simulation box the smaller is the overlap between $f$ and the bound-state wave function, and $\psi_T$ provides an accurate description of the gas-like state in the very dilute regime, but in general exhibits a non-zero overlap with the state where dimers are formed \cite{Note2}.

\begin{figure}
\begin{center}
\includegraphics[width=7.0cm]{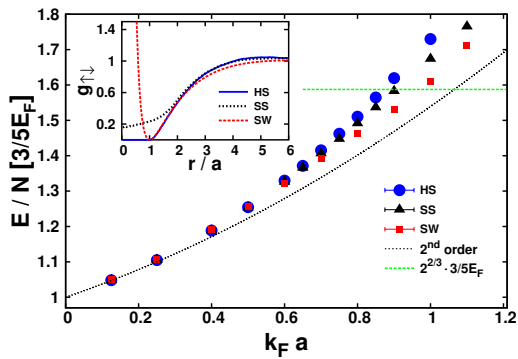}
\caption{(color online). Equation of state of the unpolarized gas. We show FN-DMC results for the repulsive HS and SS potentials and VMC results for the attractive SW potential. The perturbation expansion Eq.~(\ref{perturbation1}) is shown with the (black) dashed line, while the (green) horizontal line corresponds to the energy of the fully ferromagnetic state. Inset: Pair correlation function $g_{\uparrow\downarrow}(r)$. The range of the SS and SW potential is respectively $R_0=2a$ and $R_0=0.06a$. The minimum  at $r=a$ for the SW potential corresponds to the node in the Jastrow correlation term $f(r)$.}
\label{fig1}
\end{center}
\end{figure}

It is important to note that the repulsive HS and SS potential and the attractive SW potential, even though they correspond to the same value of the scattering length $a$ and consequently share a similar long-range behavior of the correlation functions, exhibit completely different short-range correlations as explicitly shown in the inset of Fig.~\ref{fig1} where we report results on the antiparallel spin pair correlation function.  

Simulations are performed with a maximum number of atoms in a single spin component $N_{\uparrow(\downarrow)}=33$.  For most results we checked the influence of finite-size effects by repeating the calculations with $N_{\uparrow(\downarrow)}=81$. For the results of the SW potential, no appreciable change is found by reducing the potential range parameter $nR_0^3$.  

The equation of state of the repulsive dilute gas of fermions with balanced populations, $N_\uparrow=N_\downarrow$, has previously been calculated using perturbation theory to second order in $a$: \cite{Perturbation}
\begin{equation}
\frac{E}{N}=\frac{3}{5}E_F\left[ 1+\frac{10}{9\pi}k_Fa+\frac{4(11-2\log2)}{21\pi^2} (k_Fa)^2\right] \;,
\label{perturbation1}
\end{equation}
where $E_F=\hbar^2k_F^2/2m$ is the Fermi energy. One should notice that higher order terms in the above equation will depend not only on the scattering length $a$, but also on other details of the interatomic potential. As can be seen in Fig.~\ref{fig1}. agreement between the QMC results and Eq. (\ref{perturbation1}) is found for $k_Fa\lesssim 0.4$, but significant deviations and a gradual loss of universality become evident for larger values of the interaction strength. In the figure we also show the energy of the fully ferromagnetic (FF) state $E_{\text{FF}}=3/5E_F2^{2/3}(N_\uparrow+N_\downarrow)$ consisting of two spatially separated regions of non-interacting spin-up and spin-down fermions. Any uniform mixture of the two spin species whose energy exceeds the value $E_{\text{FF}}$  is clearly unstable against phase separation.

\begin{figure}
\begin{center}
\includegraphics[width=7.0cm]{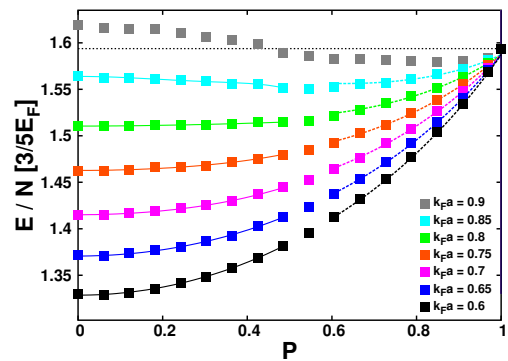}
\caption{Energy of the HS gas as a function of polarization for different values of $k_Fa$. The symbols correspond to FN-DMC results, while the lines at small and large $P$ correspond, respectively, to the fitted quadratic law and polaron energy functional of Eq.~(\ref{Hpolaron}). The horizontal dotted line is the threshold energy $E_{FF}$ of the fully ferromagnetic state.}
\label{fig2}
\end{center}
\end{figure}

In order to better characterize the critical behavior at the onset of ferromagnetic behavior, we calculate the equation of state of the gas as a function of the system polarization $P=(N_\uparrow-N_\downarrow)/(N_\uparrow+N_\downarrow)$ and then show that for $P\lesssim1$ it can be well described in terms of weakly interacting polarons. Results for the HS potential are shown in Fig.~\ref{fig2}, and similar calculations were carried out also for the SW potential (not shown). Finite size effects are reduced by subtracting from the QMC results, the finite size corrections to the ground state energy $E_0(N_\uparrow,N_\downarrow)-E_0^{\text{TL}}(P)$ of non-interacting fermions with the same number of particles and the same polarization $P$ (TL refers here to the thermodynamic limit). The validity of this method, that relies on Fermi liquid theory, is discussed in Ref.~\cite{Lin01} where it is compared with twist-averaged boundary conditions.

\begin{figure}[b]
\begin{center}
\includegraphics[width=7.0cm]{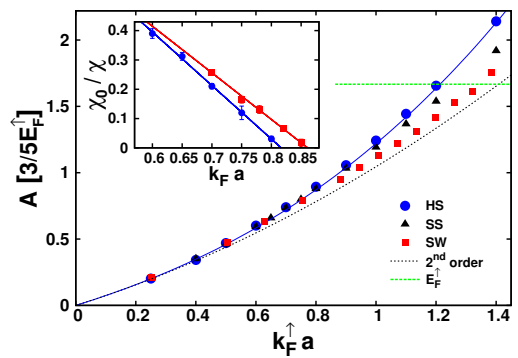}
\caption{(color online). Chemical potential at zero concentration of the repulsive polaron. We show FN-DMC results for the repulsive HS and SS potentials and VMC results for the attractive SW potential. The (blue) solid line is a fit to the HS results. The second-order perturbation expansion is shown with the (black) dashed line, while the (green) horizontal line corresponds to $E_{F\uparrow}$. Inset: Inverse magnetic susceptibility $1/\chi$ for the HS (blue circles) and SW (red squares) potential. The lines are linear fits to the data.}
\label{fig3}
\end{center}
\end{figure}

From the results at small polarization we extract the inverse magnetic susceptibility $1/\chi$ using a quadratic fit: $E(P) = E(P=0) + NE_F (\chi_0/\chi) P^2 / 3$, where $\chi_0=3n/2E_F$ is the susceptibility of the non-interacting gas.  The results for both the HS and SW potential are shown in the inset of Fig.~\ref{fig3}. For $k_Fa>0.6$ we find a linear decrease of the inverse magnetic susceptibility with increasing interaction strength, showing that $\chi$ diverges at the critical densities $k_Fa\simeq0.82$ and $k_Fa\simeq0.86$, respectively for the HS and the SW potential. As seen in  Fig.~\ref{fig2} a minimum in $E(P)$  at finite polarization $\bar{P}$ appears when $k_Fa>0.82$. This minimum corresponds to a thermodynamically stable mixed state with partially polarized domains (see Fig.~\ref{fig4}). For $P<\bar{P}$ the system follows the coexistence line where, in the thermodynamic limit, $E(P)$ is the sum of energies of domains with polarization $\bar{P}$ and $-\bar{P}$ whose relative volume changes with $P$.  For $P>\bar{P}$, on the other hand, a uniform mixture of spin components can still survive. 

For large polarizations $P$ the QMC results are well fitted by the energy functional of weakly interacting polarons 
\begin{equation}
E(x) = \frac{3}{5}N_\uparrow E_{F\uparrow}\left(1+Ax+\frac{m}{m^\ast}x^{5/3}+Fx^2\right) \;,
\label{Hpolaron}
\end{equation}  
which results from an expansion in the small concentration $x=N_\downarrow/N_\uparrow$ of spin-down impurities in a Fermi sea of spin-up particles. Here $E_{F\uparrow}=\hbar^2k_{F\uparrow}^2/2m$ is the Fermi energy of the spin-up particles in terms of the corresponding wave vector $k_{F\uparrow}=(6\pi^2n_\uparrow)^{1/3}$. Furthermore, the quantities $A$, $m^\ast$ and $F$, which are all functions of $k_{F\uparrow}a$, denote respectively the chemical potential at zero concentration, the effective mass and the interaction parameter of the polaronic quasiparticles (see Ref.~\cite{Polaron1} for the similar treatment of attractive polarons). 

The results for $A$ as a function of the interaction strength are shown in Fig.~\ref{fig3}. The weak coupling result $A=E_{F\uparrow}(4k_{F\uparrow}a/3\pi+2(k_{F\uparrow}a)^2/\pi^2)$, calculated to second order in the scattering length \cite{Polaron3,Cui10}, is also shown. This quantity is important in order to understand the stability of the fully ferromagnetic phase: in fact, if $A<E_{F\uparrow}$, the fully polarized domain is unstable against mixing of spin-up and spin-down particles resulting in a partially polarized state. This problem was investigated using a single particle-hole wave function in Ref.~\cite{Cui10}. We estimate  that  the FF state becomes unstable at $k_{F\uparrow}a=1.2$ and $k_{F\uparrow}a=1.4$, respectively for the HS and SW potential. By imposing pressure and chemical potential equilibrium between highly polarized states described by the energy functional (\ref{Hpolaron}), we determine the critical polarization boundaries, shown in the phase diagram of Fig.~\ref{fig4}, separating the homogeneous unbalanced mixture from the partially ferromagnetic state comprised of domains with opposite critical polarization $\pm\bar{P}$. This determination of the phase boundaries is reliable at large total polarization $P$, where the energy functional (\ref{Hpolaron}) is valid,  but becomes less accurate at intermediate values of $P$ (dashed line in Fig.~\ref{fig4}) and can not be applied at small polarization. In the regime of large and intermediate $P$ we find good agreement with the critical polarization as determined from the minimum of the $E(P)$ curve. The critical polarization has a sharp drop to $P=0$ close to the density where $\chi$ diverges, not allowing for a clear distinction of the order of the phase transition. The possible emergence of new phases, such as the spin textured phase proposed in Ref.~\cite{Conduit09}, also requires more detailed investigations. 

In conclusion we have calculated the equation of state of a repulsive gas of fermions as a function of interaction strength and spin polarization. We have determined the critical density for the onset of ferromagnetic behavior and we have investigated the stability of the fully ferromagnetic state. We find, however, that while the qualitative features of the phase diagrams are well described by just the long-range repulsive correlations induced by the positive $s$-wave scattering length, the quantitative determination of the phase diagram depends on the details of the interaction potential.

We acknowledge useful discussions with A. Recati. This work was supported by the Swiss National Science Foundation and by a grant from the Army Research Office with funding from the DARPA OLE program. As part of the European Science Foundation EUROCORES Program ``EuroQUAM-FerMix'' it was supported by funds from the CNR and the EC Sixth Framework Programme.

\end{document}